\documentclass[acmconf,authorversion]{acmart}

\AtBeginDocument{%
  }

\setcopyright{acmlicensed}
\copyrightyear{2024}
\acmYear{2024}
\acmDOI{10.48550/arXiv.2405.02338}

\acmConference[2024 ACM SIGGRAPH DAC Online Exhibition]{The Future of Reality: Post-Truths, Digital Twins, and Doppelgängers}{31 July 2024}{Denver, CO, USA}

\usepackage{subcaption}
\usepackage{graphicx}

\begin{document}
\title[Surrealism Me]{\textit{Surrealism Me}: Interactive Virtual Embodying Experiences in Mixed Reality}

\titlenote{This manuscript is a more formal description of the artwork Surrealism Me shown at the 2024 ACM SIGGRAPH DAC Online Exhibition -  The Future of Reality: Post-Truths, Digital Twins, and Doppelgängers. More details can be found at: https://dac.siggraph.org/artwork/surrealism-me/}

\author{Aven-Le ZHOU}
\email{aven.le.zhou@gmail.com}
\orcid{0000-0002-8726-6797}
\affiliation{%
  \institution{The Hong Kong University of Science and Technology (Guangzhou)}
  \streetaddress{No.1 Du Xue Rd, Nansha District}
  \city{Guangzhou}
  \state{Guangdong}
  \country{P.R.China}
}

\begin{teaserfigure}
  \centering
  \includegraphics[width=\textwidth]{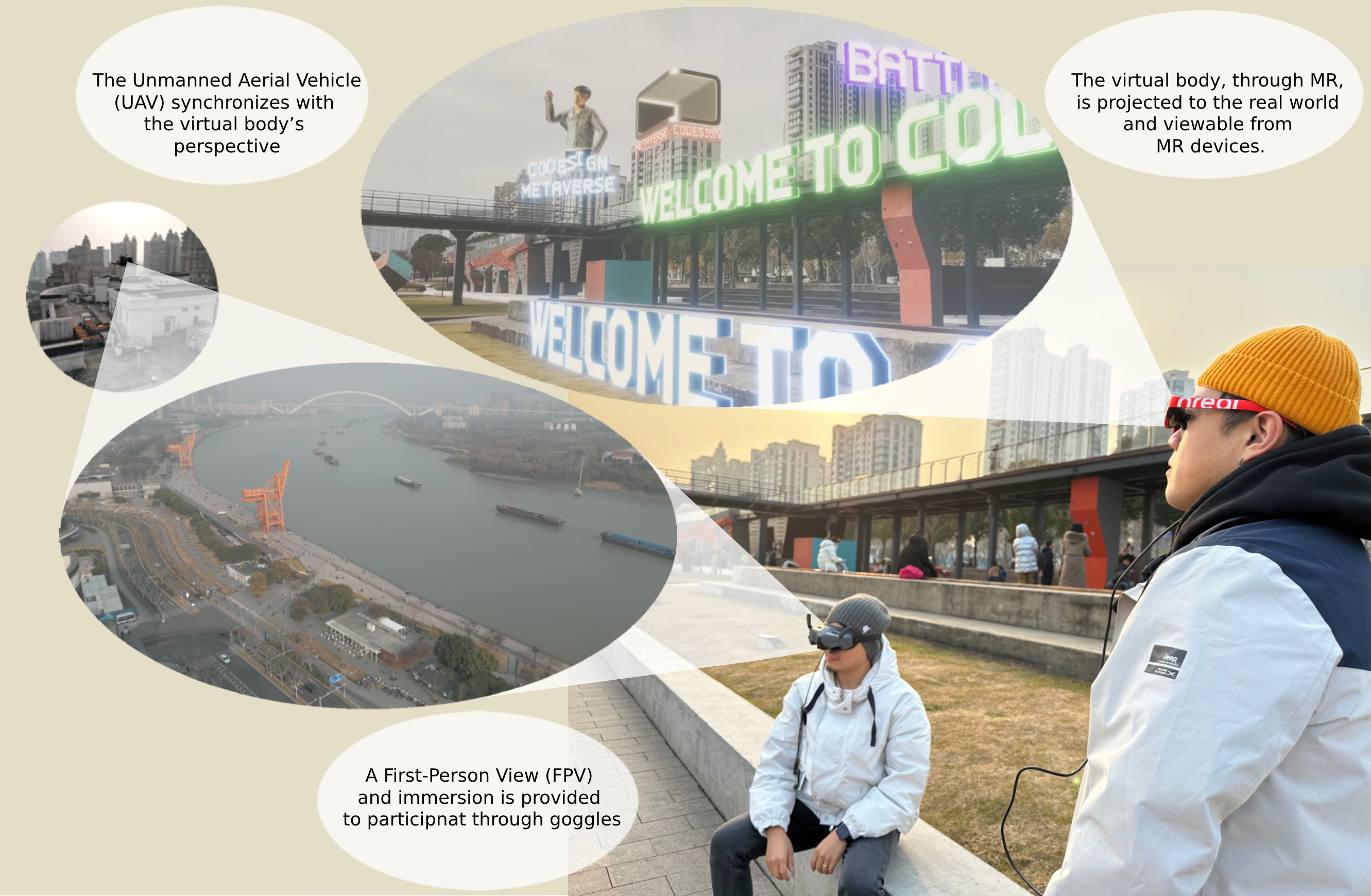}
  \caption{Interactive MR experience provided by ``Surrealism Me:'' The participant’s virtual body, through MR, is projected to the real world and viewable from MR devices. An Unmanned Aerial Vehicle (UAV) synchronizes with the virtual body’s perspective, and a First-Person View (FPV) with real-time video immersion is provided to the participant through goggles.}
  \Description{Interactive MR experience provided by Surrealism Me: The participant’s virtual body, through MR, is projected to the real world and viewable from MR devices. An Unmanned Aerial Vehicle (UAV) synchronizes with the virtual body’s perspective; and a First-Person View (FPV) with real-time video immersion is provided to participant through goggles.}
  \label{fig:teaser}
\end{teaserfigure}

\renewcommand{\shortauthors}{Aven Le Zhou.}

\begin{abstract}
This paper introduces an interactive Mixed Reality (MR) experience and artistic inquiry entitled ``Surrealism Me'' which delves into Vilém Flusser’s critique of media as mediators that often distort human perception of reality and diminish freedom, particularly within the context of MR technology. It engages with Flusser’s theories by allowing participants to experience a two-phase virtual embodying  (i.e., another body) in MR, highlighting the complex interplay between human agency, body ownership, and self-location. Initially, the participant manipulates their virtual body through various inputs or chooses AI-generated movements. Then, the interactive MR experience leads to an immersive phase where an Unmanned Aerial Vehicle (UAV) extends their sensory perceptions, embodying the virtual body’s perspective. ``Surrealism Me'' confronts the concept of ``playing against the apparatus'' by offering an interactive milieu where human and AI collaboratively explore the program's capacity limitation, thereby challenging and exhausting the potential of technology. This process further critically examines the obfuscating nature of media; as the MR medium breaks down, the work reveals the constructed nature of media-projected realities, prompting a reevaluation of media's role and influence on our perception. By navigating the boundary between real and virtual, ``Surrealism Me'' fosters a critical discourse on media’s dominance and advocates for a nuanced understanding of Flusserian freedom, encouraging participants to question and reflect on the authentic and mediated experiences of reality.
\end{abstract}




% \received{20 February 2007}
% \received[revised]{12 March 2009}
% \received[accepted]{5 June 2009}

\maketitle

\section{Contextualization}
Czech-Brazilian philosopher Vilém Flusser’s theories on media suggest that they act as mediators between humans and the world, often leading to a misperception of reality and a consequent loss of freedom \cite[p.~9]{flusser2000} \cite[p.~2]{flusser2013}. In the context of Mixed Reality (MR), this mediation can obscure the world further, complicating our relationship with the MR media. This paper introduces ``Surrealism Me'' to address this concept by offering an interactive art experience that allows participants to virtually embody their presence, following the “Sense of Embodiment” \cite{6797786} definitions in MR.

\section{Sense of Embodiment}

% \begin{figure}[h]
%   \includegraphics[width=\textwidth]{bodily-movement.jpg}
%   \caption{SoE.2023}
%   \Description{Fig.1. Interactive MR Experience Provided by Surrealism Me.}
%   \label{fig:teaser}
% \end{figure}

\begin{figure}[h]
  \centering
  \begin{subfigure}{0.475\textwidth}
      \includegraphics[width=\textwidth]{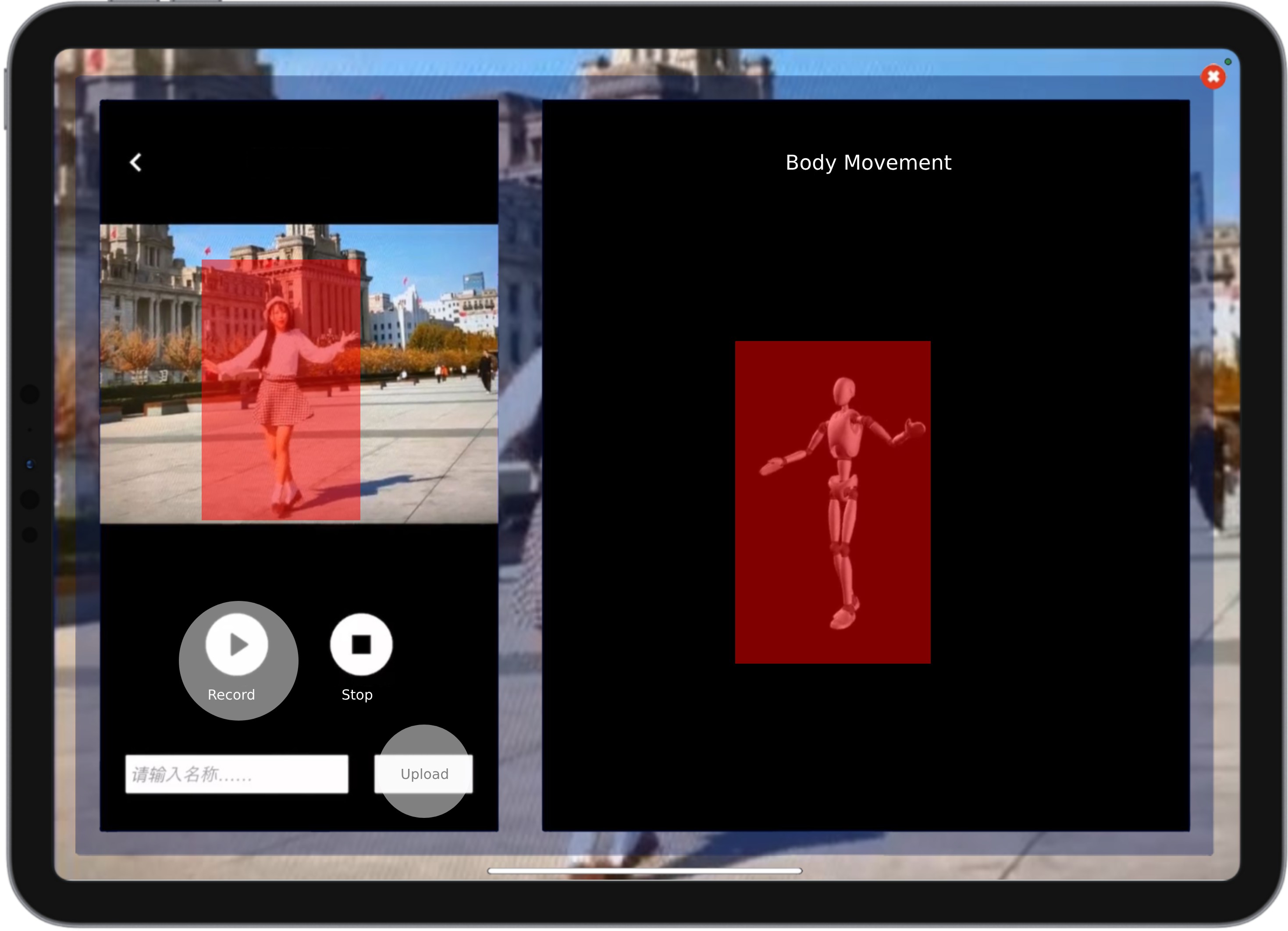}
      \caption{Participants moving their body in front of the camera, and recorded motion controls the virtual body.}
      \label{fig:motion-left}
  \end{subfigure}
  \hfill
  \begin{subfigure}{0.475\textwidth}
      \includegraphics[width=\textwidth]{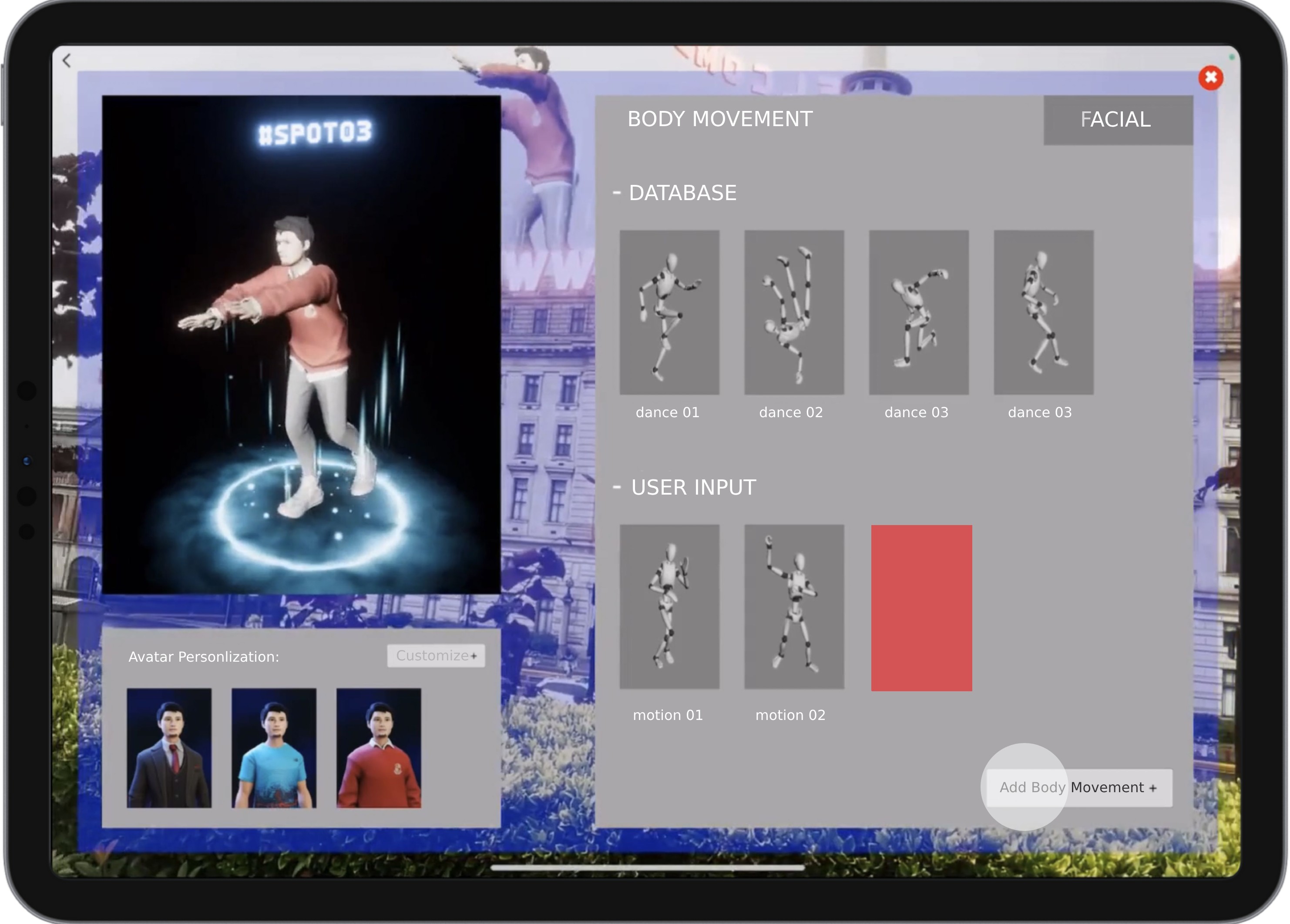}
      \caption{Interface to choose the preferred motion of their recorded movements or select AI-generated ones.}
      \label{fig:motion-right}
  \end{subfigure}
  \caption{Being the author of virtual body’s movements: Participants controlling their virtual body.}
  \label{fig:motion}
  \end{figure}

What does it feel like to own, to control, and to be inside a virtual body? Minsky, when coining the term ``Telepresence,'' suggested that the feeling of presence and the sensation of being there are key challenges to achieve \cite{minsky1980telepresence}. Kilteni and Slater’s framework further defines the immersion and presence in a virtual body through three aspects: (1) Agency, the sense of being the author of a body’s movements; (2) Body Ownership, the feeling that the body is the source of experienced sensations; and (3) Self-Location, the sense of being located inside the body \cite{6797786}. This multidimensional experience, coupled with the continuous presence of one’s biological body, raises questions about the extent to which one can experience the same sensations towards a virtual body as toward the biological body.

% \begin{figure}[h]
%   \centering
%   \includegraphics[width=0.475\textwidth]{attached.jpg}
%   \caption{Virtual content anchored in physical world as MR experience.}
%   \label{fig:MRtech-c}
%   \end{figure}

  % \begin{figure}[h]
  % \centering
  % \includegraphics[width=0.475\textwidth]{surrealismme-main.jpg}
  % \caption{Interactive MR experience provided by ``Surrealism Me:'' The participant’s virtual body, through MR, is projected to the real world and viewable from MR devices. An Unmanned Aerial Vehicle (UAV) synchronizes with the virtual body’s perspective, and a First-Person View (FPV) with real-time video immersion is provided to the participant through goggles.}
  % \label{fig:MRtech-c}
  % \end{figure}

\section{Interactive Art Experience}

% \begin{figure}[h]
%   \includegraphics[width=0.95\textwidth]{bodily-movement.jpg}
%   \caption{Being the author of virtual body’s movements: Participants controlling their virtual body.}
%   \Description{Fig.1. Interactive MR Experience Provided by Surrealism Me.}
%   \label{fig:place}
% \end{figure}

``Surrealism Me'' enables participants to create a personalized avatar for a comprehensive virtual embodying experience in MR. This process involves two phases; initially, participants can manipulate their virtual body in real time by controlling its movements through various inputs. Participants can move their bodies in front of the camera, record, and upload the footage representing their physical movements (as in Fig. \ref{fig:motion-left}) or select AI-generated motions as their preferred body movements (see Fig. \ref{fig:motion-right}). This level of control enables users to author the movements of their virtual body, as the first layer of virtual embodying experience. The virtual body, through MR, is projected to the real world and viewable from MR devices.

\begin{figure*}[h]
  \includegraphics[width=\textwidth]{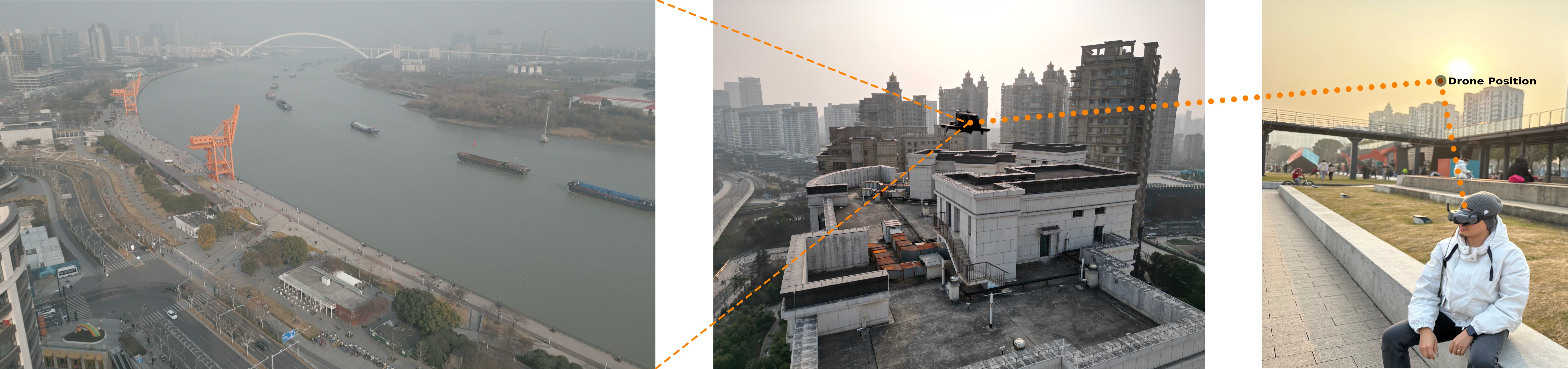}
  \caption{Experiencing sensations to feel being located inside the virtual body.}
  \Description{Fig.1. Interactive MR Experience Provided by Surrealism Me.}
  \label{fig:vision}
\end{figure*}

In the second phase, ``Surrealism Me'' employs an Unmanned Aerial Vehicle (UAV) that serves as the visual and auditory sensor of the user’s virtual body. The UAV's position and perspective is synchronized with the virtual body; the real-time video is provided to the participant through a first-person view (FPV) goggles as the real-time immersive experience. Participants then experience visual and auditory sensations, allowing them to feel locating and immersing themselves within their virtual body.

\section{Summarization }

In its first phase, ``Surrealism Me'' offers an opportunity to engage with Flusser's notion of ``playing against the apparatus'' \cite [pp.~80-82] {flusser2000} through ``exhausting the apparatus'' \cite{poltronieri2014communicology} in the context of Artificial Intelligence. Participants explore various possibilities through direct bodily interaction and other unconventional modalities. Additionally, their motion data input is archived as a training dataset. Through the AI’s learning and generative capabilities, humans and AI collaboratively exhaust the program's potential.

Moreover, ``Surrealism Me'' eventually exposes the obfuscating effects of media by demonstrating how MR media breaks down, revealing the disparity between MR and reality. The transition from the immersive experience provided by the UAV’s perspective to the recognition of a simulated environment underscores a sense of alienation, leading to the realization that the MR world, though seemingly real, is a mere projection by the apparatus. As the medium breaks down, the project reveals the constructed nature of media-projected realities and critically examines nature of MR media. ``Surrealism Me'' thus challenges participants' perceptions of media reality, offering critical reflection on the media's influence on our understanding of the world. It encourages a reevaluation of media’s role and influence on our perception as its dominance and advocates for Flusserian Freedom.

%% The next two lines define the bibliography style to be used, and
%% the bibliography file.
\bibliographystyle{ACM-Reference-Format}
\bibliography{sample-base}

\end{document}